\documentclass[12pt]{iopart}
\usepackage{iopams,epsf}

\usepackage{color}
\usepackage{epsfig}
\usepackage{graphics}
\usepackage{psfrag}
\usepackage{lscape}

\begin{document}

\title[Symmetries of microcanonical entropy surfaces]{Symmetries of microcanonical entropy surfaces}

\author{Hans Behringer}

\address{Institut f\"ur Theoretische Physik I, Universit\"at Erlangen-N\"urnberg, D -- 91058 Erlangen, Germany}

\address{Institut Laue-Langevin, F -- 38042 Grenoble, France}

\ead{Hans.Behringer@physik.uni-erlangen.de }

\begin{abstract}
Symmetry properties of the microcanonical entropy surface as a function of the energy and the 
order parameter are deduced from the invariance group of the 
Hamiltonian of the physical system. The consequences of these symmetries for the microcanonical order 
parameter in the high energy and in the low energy phases are investigated. In particular the breaking of the 
symmetry of the microcanonical entropy in the low energy regime is considered. 
The general statements are corroborated by 
investigations of various examples of classical spin systems. 
\end{abstract}



%\maketitle

\section{Introduction}

Statistical mechanics relates the macroscopic thermodynamics of physical systems to its 
microscopic properties. The starting point of the statistical description is the density of states 
or equivalently the microcanonical entropy which is usually transformed to the canonical partition function. In recent years the 
statistical properties of various systems have been deduced directly from the microcanonical 
entropy \cite{huell94,gros01}. In particular first order phase transitions 
\cite{schm94,wale94,gros96} and critical phenomena have been extensively 
analysed \cite{prom95,kast00}. Those investigations reveal that certain properties of phase transitions can already be 
observed in finite systems making the use of the microcanonical approach
advantageous. Note however that the choice of the appropriate ensemble is
strongly related to the experimental setup of the physical system. 
Further focus is laid on the structure of the entropy surface as a geometrical object \cite{Gros00,Plei01,Kast02} 
and the question of the equivalence of the different statistical ensembles \cite{Lewi94,Daux00,Barr01,Ispo01}.
Note also that apart from the microcanonical ensemble precursors of phase transitions of an infinite system in the
corresponding finite system can be investigated from different perspectives
such as the distribution of Yang-Lee zeros of the finite system partition
function \cite{borr00} or the topology properties of the configuration space
\cite{caia97}.

The topic of the present paper are the symmetries of the microcanonical entropy surface of finite spin systems. These symmetry properties 
are deduced from the symmetries of the microscopic Hamiltonian. Additionally they allow statements 
about the microcanonically defined order parameter. The order parameter in the microcanonical ensemble 
exhibits the characteristics of spontaneous symmetry breaking and its symmetry properties are 
related to those of the microcanonical entropy surface. 

The work is organised in the following manner. The second section contains a 
recapitulation of the some of the basic concepts of the 
investigation of physical systems in the microcanonical approach. In the third section the 
question of the symmetries of spin systems and of the corresponding consequences on the  
microcanonical entropy is 
discussed. The general findings are investigated for particular examples of classical spin systems in the 
fourth section. The entire treatment is formulated in the language of magnetic spin systems  
although the statements are more general and apply to other physical situations as well.

\section{Spin systems in the microcanonical ensemble}

Consider a subset $\mathcal{P}$ of $N$ points in the $d$-dimensional space $\mathbb{Z}^d$ with a local 
spin degree of freedom $\sigma_i$ at each lattice site $i$. On the microscopic level 
a physical system is defined by its Hamiltonian $\mathcal{H}$.
The Hamiltonian contains all contributions to the total energy of the system originating from all possible 
internal interactions of the spin variables $\sigma_i$. For an isolated system, 
i.e. no external field is applied to the spin 
variables, and pair interactions only the general Hamiltonian can be written as 
\begin{equation}
	\label{eq:hamilton_allgemein}
	\mathcal{H}(\sigma) = \sum_{(i,j) \in \mathcal{P}\times\mathcal{P}} c_{ij} \mathcal{I} (\sigma_i, \sigma_j) \;.
\end{equation}
The pair interaction of the two spins $\sigma_i$ and $\sigma_j$ is given by $\mathcal{I}$. Prominent examples of the 
interaction term $\mathcal{I}$ are the Ising model with $\mathcal{I}(\sigma_i, \sigma_j)= \sigma_i\sigma_j$ and the $q$-state 
Potts model with $\mathcal{I}(\sigma_i, \sigma_j)= \delta_{\sigma_i,\sigma_j}$. The Ising spins can take on the values $\sigma_i = \pm 1$ 
whereas the Potts spins can be in the states $\sigma_i = 1, \ldots, q$. The coupling constants 
$c_{ij}$ describe the strength of the spin interactions of the various sites $i$ and $j$. These constants also 
define the range of the interaction. Note that the boundary conditions of the finite subset $\mathcal{P}$ in 
$\mathbb{Z}^d$ are also specified by the constants $c_{ij}$. 
To describe the magnetic properties of a classical spin system the total magnetisation as a further macroscopic 
quantity is used. The magnetisation of a given spin configuration $\sigma$ is related to an operator $\mathcal{M}$. 
In general the magnetisation operator is a multicomponent object:
\begin{equation}
	\mathcal{M}  = (\mathcal{M}_1, \ldots, \mathcal{M}_n)
\end{equation}
with $n$ being the number of components.
For most classical spin systems the magnetisation is just the sum of the local 
spin variables 
\begin{equation}
	\mathcal{M}(\sigma) = \sum_{i \in \mathcal{P}} \sigma_i
\end{equation}
but it might also be a more complicated linear function of the spins $\sigma_i$.
For an antiferromagnetic spin model for example the operator $\mathcal{M}$ denotes the staggered magnetisation. 
In general the operator $\mathcal{M}$ describes a macroscopic quantity that allows the definition of the microcanonical 
order parameter (see relation (\ref{eq:def_spontane_mag}) below). The following sections stick to the language of magnetic systems.

The two variables energy $E$ and magnetisation $M$ specify a so-called
macrostate of the magnetic system. For a microstate $\sigma$ these two
quantities are obtained by applying the operators $\mathcal{H}$ and
$\mathcal{M}$.  
Several different microstates $\sigma$ can belong to the same 
macrostate $(E,M)$. Let $\Gamma_N$ be the phase space of all possible microstates of a physical system of 
$N$ spins on a $d$-dimensional hypercube with linear extension $L$. The density of states of a discrete spin system is defined by 
\begin{equation}
	\label{eq:def_dos}
	\Omega(E, M, L^{-1}) = \sum_{\sigma \in \Gamma_N} \delta_{E, \mathcal{H}(\sigma)}    
	\delta_{M, \mathcal{M}(\sigma)} 
\end{equation}
where the Hamiltonian $\mathcal{H}$ and the magnetisation operator $\mathcal{M}$ give 
the internal energy and the magnetisation of the microstate $\sigma$. 
For a discrete spin model with discrete values for $E$ and $M$ the density of states $\Omega$ is the number 
of different microstates which are compatible with a specified macrostate.

The density of states is the starting point of the statistical description of thermostatic 
properties of a physical system. These properties are deduced from the corresponding 
thermodynamic potential. The set of natural variables on which this potential depends is 
determined by the physical context. For a magnetic system that is isolated from any environment 
the proper natural variables are the energy $E$ and the magnetisation $M$. 
The thermodynamic potential of 
an isolated system is the microcanonical entropy 
\begin{equation}
	\label{eq:def_entropie}
	S(E, M, L^{-1}) = \ln \Omega(E,M, L^{-1}) \;.
\end{equation}
Here and in the following natural units with $k_B = 1$ are used.  

In the conventional approach the  system is coupled to an (infinitely) 
large reservoir so that it can exchange energy and magnetisation with 
its surrounding. The corresponding thermodynamic potential, the Gibbs free energy, 
is connected to the canonical partition function
\begin{equation}
	\label{eq:def_zustandssumme}
	Z(\beta, \beta h, L^{-1}) = \sum_{(E,M)} \Omega(E,M, L^{-1}) \exp(-\beta(E - hM)) 
\end{equation}
by the relation
\begin{equation}
G(\beta,\beta h, L^{-1}) = -\frac{1}{\beta}\ln Z(\beta, \beta h, L^{-1})  \;.
\end{equation}
The external magnetic field is denoted by $h$ and $\beta$ is the inverse temperature.
In the following however the microcanonical instead of the canonical approach to the thermostatic 
properties of physical systems is investigated.

The starting point of the microcanonical analysis of finite classical spin systems is the intensive 
microcanonical entropy
\begin{equation}
	s(e,m, L^{-1}) := \frac{1}{L^d} S(L^de, L^dm, L^{-1})
\end{equation}
of a system. In this quantity the trivial 
size dependence of the entropy is divided out, nevertheless $s$ will still show a non-trivial 
dependence on the system size. 
The intensive energy and magnetisation are defined by 
$e := E/L^d$ and $m := M/L^{d}$. %To improve readability this dependence of the 
%microcanonical quantities on the length $L$ is suppressed in the following. 

The spontaneous magnetisation of a finite microcanonical system with length $L$ for given energy $e$, i.e. the magnetisation in thermostatic 
equilibrium, is defined by 
\begin{equation}
	\label{eq:def_spontane_mag}
	m_{sp}(e) : \iff s(e, m_{sp}(e)) = \max_{m} s(e, m) \;.
\end{equation}
The spontaneous magnetisation of a finite system for given energy is the value $m$ at which the entropy 
becomes maximum for the fixed value of $e$, i.e. the macrostate $(e,m)$ that maximises the density 
of states defines the equilibrium macrostate. 
The non-vanishing multicomponent spontaneous magnetisation 
of the low energy phase defines a direction in the order parameter space:
\begin{equation}
	\label{eq:mag_richtung}
	m_{sp}(e) = (m_{sp}^{(1)}(e), \ldots, m_{sp}^{(n)}(e)) = 
	|m_{sp}(e)| \mu^{(0)}   
\end{equation}
where $\mu^{(0)}$ is a unit vector. 
It should be noted that 
several equivalent maxima of the entropy may exist for a given energy $e$. This 
appearance of different but equivalent equilibrium macrostates is related to spontaneous 
symmetry breaking (see below). 
The spontaneous magnetisation defines the 
order parameter. For the high energy phase the microcanonical order parameter is zero reflecting 
a phase with high symmetry. The investigation of finite spin systems within the microcanonical approach 
reveals that the microcanonically defined order parameter (\ref{eq:def_spontane_mag}) 
is indeed zero for energies above a critical value $e_c$ and 
becomes non-zero below this energy \cite{kast00}. Near the energy $e_c$ the magnetisation exhibits a 
square root dependence on 
the energy difference $e-e_c$ for all finite system sizes. 
The singular behaviour of the order parameter of a finite microcanonical 
system can therefore be characterised by a critical exponent $\tilde{\beta}= 1/2$. 
Here some comments on the used language are necessary. The magnetisation of the finite system 
is made up of discrete data points. The energy dependence of this curve for small magnetisations is 
most suitably described by a continuous square root function. The foot of this curve defines the critical 
energy $e_c$ of the finite system. %The expression {\em critical} is used due to the striking similarity 
%of the behaviour of the physical quantities of the finite 
%microcanonical systems to the physics of phase transitions in infinite systems. 
%Nevertheless one has to bear in mind that finite systems are considered whose characteristics 
%are described using the language of critical phenomena in the thermodynamic limit. 
 
The abrupt emerging of a finite order parameter at 
the critical value $e_c$ indicates the 
transition to an ordered phase with lower symmetry.  
Although the physical quantities like the order parameter or the susceptibility 
show singularities \cite{kast00} at the transition point that are typical of phase transitions 
there is no phase transition in a finite microcanonical system. The microcanonical entropy of finite 
systems is expected to be an analytic function of its natural variables and hence a phase transition 
in the narrow sense 
does not take place. However, for all the author knows the analyticity of the microcanonical entropy is not yet proven, nevertheless it 
seems natural to assume this property regarding the analyticity of the canonical potential of 
finite systems. The use of the expressions {\em phase transition} 
and {\em critical} to describe 
properties of theories that account for singular behaviours in 
physical quantities but that are based on analytic potentials is familiar in the context of molecular field 
approximations.

\section{Symmetry properties of the entropy}

A physical system specified on a microscopic level by its Hamiltonian 
$\mathcal{H}$ may exhibit certain symmetries.\footnote{Symmetry properties of classical spin systems are 
treated in the textbook \cite{mora01}.} 
A symmetry transformation $g$ 
is defined by the fact that it leaves the interaction energy of all configurations $\sigma$ of the phase space $\Gamma_N$  invariant, 
i.e. for all microstates one has 
\begin{equation}
	\mathcal{H}(\sigma) = \mathcal{H}(g(\sigma)) \;.
\end{equation}
One 
distinguishes between space symmetries and internal isospin symmetries. In the following the group 
$G_L$ of space symmetries of the lattice $\mathcal{P} \subset \mathbb{Z}^d$ is not further regarded. 
The Hamiltonian may also be invariant under transformations that act solely on the internal degrees of freedom $\sigma_i$. 
Such a transformation $g_S$ maps the spin variable $\sigma_i$ onto the new spin variable $g_S(\sigma_i)$. The map $g_S$ transforms 
the spins of different lattice sites in the same way constituting therefore a global internal symmetry. 
The set of these internal symmetry 
transformations defines the isospin group $G_S$ of the spin model. The total symmetry group of the Hamiltonian $\mathcal{H}$ is then 
given as the direct product $G = G_L \otimes G_S$. Although the Hamiltonian is invariant under the transformations of $G_S$, 
the magnetisation $\mathcal{M}(\sigma)$ need not be identical to the magnetisation $\mathcal{M}(g_S(\sigma))$ of the transformed 
configuration $g_S(\sigma)$. An element $g_S$ of the
group $G_S$ induces a transformation
\begin{equation}
	\label{eq:darstellung0}
	\mathcal{M}(\sigma) \stackrel{g_S}{\longmapsto}  \mathcal{M}(g_s(\sigma))
\end{equation}
on the magnetisation space spanned by the components $\mathcal{M}_l(\sigma)$ with $l = 1, \ldots, n$.
For physically
relevant magnetisation operators 
$\mathcal{M}$ (see the discussed examples in Sec. \ref{kap:beispiele}) this induced
transformation defines a $n$-dimensional representation
$D(G_S)$ of the group $G_S$ by the correspondence 
\begin{equation}
g_S \longmapsto D(g_S)
\end{equation}
with $D(g_S)$ denoting the induced map.
The transformed magnetisations $\mathcal{M}(g_S(\sigma_1))$ and
$\mathcal{M}(g_S(\sigma_2))$ of any two different configurations $\sigma_1$ and
$\sigma_2$ of the same macrostate $M$ are then  identical and thus
$g_S$ does induce unambiguously the map $D(g_S)$ onto the magnetisation space.
Note however that for an arbitrarily chosen magnetisation operator
$\mathcal{M}$ the transformed magnetisations of two different configurations $\sigma_1$ and
$\sigma_2$ of  a given  macrostate $M$ may not be identical. In such a
situation the transformation (\ref{eq:darstellung0}) does not induce a
representation of the invariance group $G_S$ as the map $D(g_S)$ is not
defined unambiguously for all $g_S$ in $G_S$. Nevertheless there might be a
non-trivial subgroup $\tilde{G}_S$ for which the transformation
(\ref{eq:darstellung0}) induces a representation $D(\tilde{G}_S)$. In the
following however it is assumed that the representation comprises the whole
invariance group $G_S$.

%However there exists always a subgroup $\tilde{G}_S$ in $G_S$ such that for
%any macrostate $M$ any two of its configurations are mapped onto the same
%magnetisation under the action (\ref{eq:darstellung0}) of an arbitrary  element
%$\tilde{g}_S \in \tilde{G}_S$. For this subgroup $\tilde{G}_S$ the
%magnetisation operator $\mathcal{M}$ induces a $n$-dimensional representation
%$D(\tilde{G}_S)$ of the subgroup $\tilde{G}_S$ by the correspondence 
%\begin{equation}
%\tilde{g}_S \longmapsto D(\tilde{g}_S)
%\end{equation}
%with $D(\tilde{g}_S)$ denoting the induced map. Note that the
%subgroup $\tilde{G}_S$ depends on the choice of the magnetisation operator
%and 
%might even be the trivial subgroup $\{1\}$. For physically
%relevant magnetisation operators however $\tilde{G}_S$ is often identical to
%$G_S$ (see the discussed examples in Sec. \ref{kap:beispiele}). For the
%following treatment it is therefore assumed that $\tilde{G}_S = G_S$. 

%This mapping defines a $n$-dimensional representation 
%$D(G_S)$ of the group $G_S$ on the $n$-dimensional magnetisation space by the correspondence
%\begin{equation}
%	g_s \longmapsto D(g_s) \;.
%\end{equation}

The symmetry property of the Hamiltonian is reflected by 
a symmetry of the microcanonical entropy as a function of the magnetisation components. 
Applying the symmetry transformation $g_S$ the macrostate $(e, m)$ is then mapped 
onto the macrostate $(e, D(g_S)(m))$ and a microstate $\sigma$ which is compatible with the macrostate 
$(e, m)$ is mapped onto the new configuration $g_S(\sigma)$. This transformed configuration is compatible 
with the transformed macrostate $(e, D(g_S)(m))$. This is obvious from the
above discussion of the induced representation. Hence any microstate belonging to the original 
macrostate is mapped onto a configuration of the new macrostate. On the other hand consider a 
configuration $\sigma^{\prime}$ of the new macrostate $(e, D(g_S)(m))$. The configuration 
$g_S^{-1}(\sigma^{\prime})$ has magnetisation $D^{-1}(g_S)(D(g_S)(m)) = m$ and contributes therefore to the 
macrostate $(e, m)$. Thus there are as many configurations belonging to the macrostate $(e, m)$ as there are 
microstates compatible with the macrostate $(e, D(g_S)(m))$. This results in the symmetry property 
\begin{equation}
	\label{eq:entropie_sym_allg}
	s(e, D(G_S)(m)) = s(e, m)
\end{equation}
of the microcanonical entropy surface.

In the following it is assumed that the group $G_S$ is finite and that the 
representation $D(G_S)$ is irreducible. In a physical situation where the 
representation $D(G_S)$ is reducible it is always possible to decompose it into its irreducible contributions. The different 
irreducible magnetisation components can be considered separately as the physical behaviours associated with them 
is independent from each other.

Consider the equilibrium macrostates of a spin system. Above a certain energy $e_c$ the system is in a phase 
with zero spontaneous magnetisation. Below the critical energy $e_c$ the system is in a stable phase with 
a non-zero microcanonical order parameter. The high energy phase corresponds therefore to the macrostate $(e,0)$ that is 
invariant under all transformations of $D(G_S)$.
The high energy phase possesses the same symmetry group $G_S$ as the Hamiltonian $\mathcal{H}$, i.e. 
the order parameter $m_{sp}(e)=0$ of the high energy phase is left invariant under the action of 
$G_S$ on the order parameter space. 
The low energy phase below $e_c$ has a non-zero equilibrium magnetisation $m_{sp}$. Then some map 
$D(g_S)$ has to transform $m_{sp}$ onto a magnetisation $D(g_S)(m_{sp})$ that is not identical to 
$m_{sp}$. If this was not the case the representation would not be irreducible. The low energy 
phase has consequently a smaller symmetry $G$ than the high energy phase. The group $G$ is a subgroup of 
$G_S$ whose representation $D(G) \subset D(G_S)$ leaves the non-zero spontaneous magnetisation 
$m_{sp}$ invariant, i.e.
\begin{equation}
	D(G)(m_{sp}) = m_{sp} \;. 
\end{equation}
In view of the decomposition (\ref{eq:mag_richtung}) of the $n$-dimensional spontaneous 
magnetisation $m_{sp}(e)$ into an energy dependent 
modulus $|m_{sp}(e)|$ (the actual order parameter) and a fixed direction 
$\mu^{(0)}$ in the order parameter space this property 
means that the direction $\mu^{(0)}$ is left invariant by the representation $D(G)$.
The symmetry of the high energy phase $G_S$ is spontaneously broken down to the symmetry $G$ of the 
low energy phase characterised by a non-zero order parameter.\footnote{Group theoretical aspects of spontaneous 
symmetry breaking are presented in \cite{Mich80,mora01}.}

The group $G_S$ can be decomposed into left cosets with respect to the subgroup $G$:
\begin{equation}
	\label{eq:nebenklassen_zer} 
	G_S = G \cup h_1G \cup \ldots \cup h_{u-1}G  \;.
\end{equation}
This decomposition is unique in the sense that all $h_iG$ are disjoint well defined sets and the number $u$ of cosets 
appearing in the decomposition (\ref{eq:nebenklassen_zer}) is given by the index of the group $G$ in 
$G_S$ \cite{lyub60,fali66}. The elements $h_i$ labelling the distinct cosets in (\ref{eq:nebenklassen_zer}) do not belong to 
$G$ and are not unambiguous as any element 
in $h_iG$ can be used as a label. The transformation $D(h_i)$ maps the unit vector $\mu^{(0)}$ onto the new direction 
\begin{equation}
	\label{eq:transformiertesym}
	\mu^{(i)} = D(h_i)(\mu^{(0)}) \neq \mu^{(0)} \;.
\end{equation}
Any other element $h_i^{\prime}$ in $h_iG$ transforms $\mu^{(0)}$ also onto the direction $\mu^{(i)}$. As all 
macrostates that are obtained from the state $|m_{sp}(e)| \mu^{(0)}$ through a transformation from the 
set $D(G_S)$ have the same value of the entropy, i.e.
\begin{equation}
	s(e, D(G_S)(|m_{sp}(e)| \mu^{(0)})) = s(e, |m_{sp}(e)| \mu^{(0)}) \;,
\end{equation}
all distinct macrostates $\{\left. |m_{sp}(e)| \mu^{(i)} \right| i = 0, \ldots, u-1 \}$ are equivalent stable 
phases below the critical energy $e_c$. Hence there are $u$ physically equivalent phases below the transition energy. 
The symmetry group $G^{(i)}$ of the phase $|m_{sp}(e)| \mu^{(i)}$ is obtained from the invariance group 
$G$ by a conjugation with the element $h_i$:
\begin{equation}
	\label{eq:konjugiertesym}
	G^{(i)} = h_i G h_i^{-1} \;.
\end{equation}

\begin{figure}
\begin{center} 
\epsfig{file=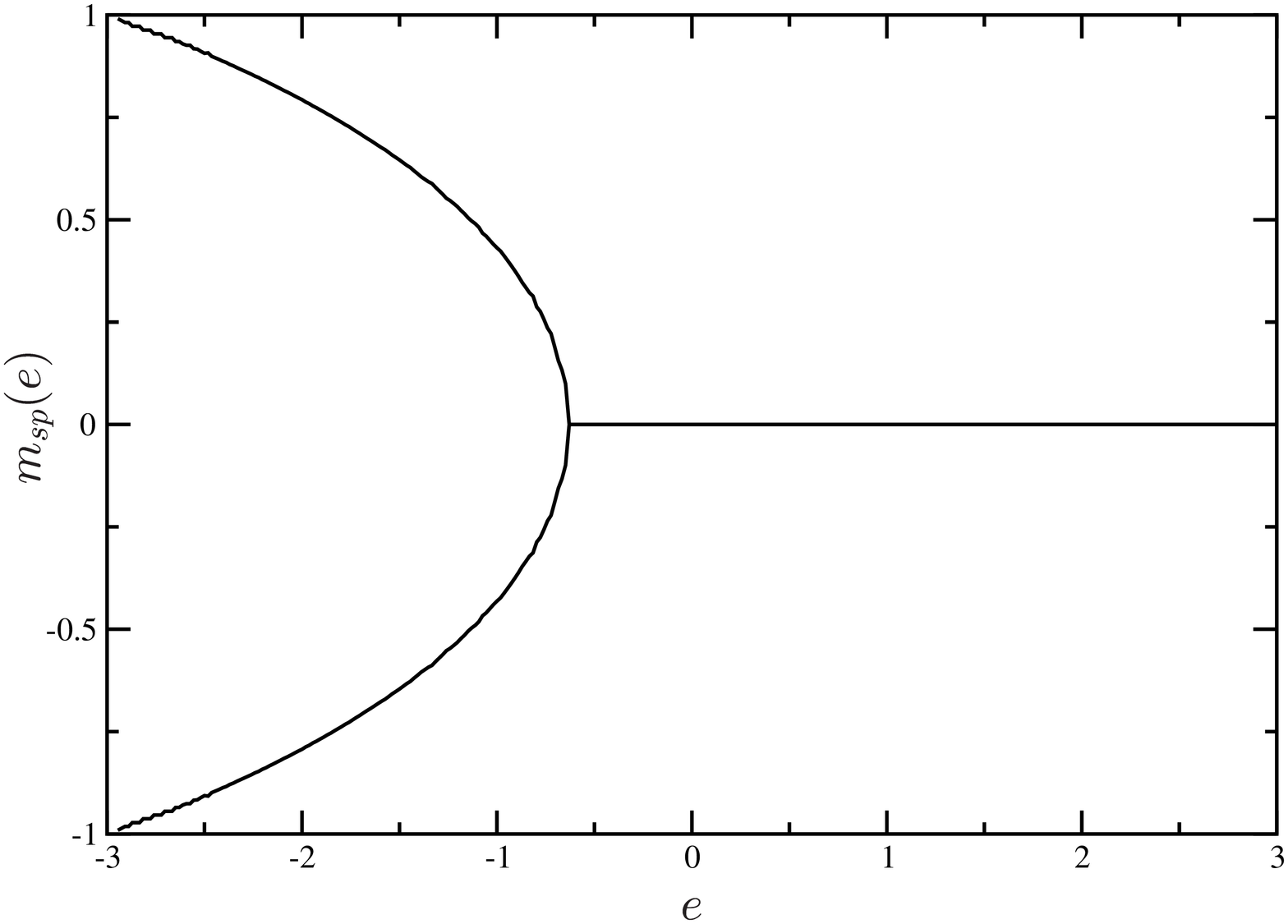, width=15em, angle=270}
\end{center}
\caption{\label{bildsponmag_ising} \small The two branches of the microcanonical order parameter 
of the three-dimensional Ising model 
with $216$ spins as a function of the energy. 
The data that are obtained by a Monte Carlo simulation suggest a critical energy $e_c = -0.665\pm 0.002$.} 
\end{figure}

To conclude this section the order parameter symmetry of the Ising model is shortly considered. The 
Hamiltonian 
\begin{equation}
	\mathcal{H} = -\sum_{\left< i,j \right>}\sigma_i \sigma_j
\end{equation}
of the nearest neighbour Ising model (the summation over neighbour pairs is indicated by 
$\left< i,j \right>$) with possible spin states $\sigma_i = \pm 1$ is invariant under 
the Abelian group $G_S= C_2 = \{ +1, -1\}$. The 
entropy as a function of the one-dimensional order parameter $m$ is consequently an even function
\begin{equation}
	s(e,-m) = s(e, m) \;.
\end{equation}
The equilibrium order parameter in the low energy phase is non-zero and hence the low energy phase is 
only invariant under the trivial subgroup $G=\{ +1\}$. The two possible order parameter branches 
are related to each other by the group transformation $-1$. 
They are shown in Fig. \ref{bildsponmag_ising} for a three-dimensional Ising system with 216 spins.

\section{Entropy surface of $q$-state models}
\label{kap:beispiele}

In this section the symmetry properties of the microcanonical entropy surfaces of the 
three-state Potts model and the four-state vector Potts model are investigated. 
Both models are examples of systems with a two-dimensional 
order parameter. The entropy is consequently a function of the 
internal interaction energy $e$ and the two order parameter 
components $m_1$ and $m_2$. For simplicity both 
models are defined  on a two-dimensional square lattice with 
linear extension $L$ and $N=L^2$ lattice sites. The boundary 
condition is chosen to be periodic. The entropy surface is obtained by the highly 
efficient transition observable method \cite{huell02} that 
allows the determination of very accurate estimates of the entropy surface.

\subsection{Three-state Potts model}

The Potts model \cite{Wu82} is a possible generalisation of the Ising model. The Hamiltonian is given by 
\begin{equation}
	\label{eq:potts_hamilton}
  \mathcal{H} = -\sum_{\left< i,j \right>} \delta_{\sigma_i, \sigma_j} \;,
\end{equation}
where the Potts spins $\sigma_i$ can take on the values $1, \ldots, q$. The case $q=2$ 
is equivalent to the Ising model. In two dimensions the model undergoes a second order 
phase transition for $q<5$.

\begin{figure}%[htb]
\begin{center}
\epsfig{file=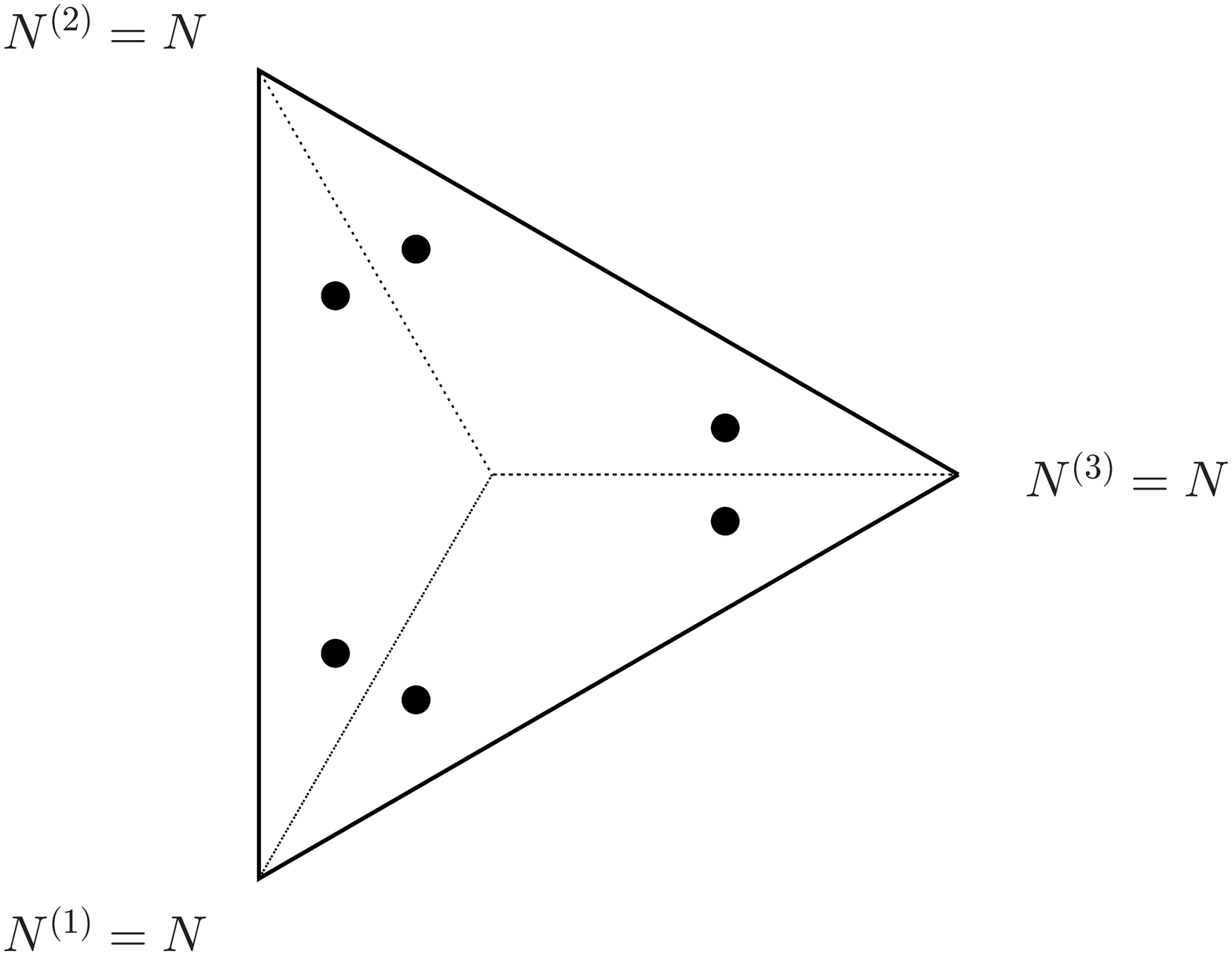, width=15em, angle=270}
\end{center}
\caption{\label{pottsdreieck} \small The equilateral triangle
spanned by the magnetisations of the three
different ground states of the three-state Potts model. The circles indicate
macrostates with the same
value of the entropy (compare relation (\ref{eq:potts_entro_sym})).}
\end{figure}

The three-state Potts model where the spin variables $\sigma_i$ can take on the values 1, 2, and 3,   
has a three-fold degenerate completely ordered ground state with internal energy 
$E = -2N$. In one of the three ground states all $N$ spin variables are in the same state. 
The energy $E$ of the configurations 
$\sigma$ can take on any integer value 
in the interval $[-2N,0]$. The magnetisation 
of the system for a fixed energy is related to the numbers $N^{(\sigma_i)}$ of spins being in the spin state 
$\sigma_i$. As the total number of spins is fixed to be $N$ these numbers are 
subjected to the subsidiary condition $N^{(1)} + N^{(2)} + N^{(3)} = N$ which defines a plane 
in the three-dimensional space spanned by the $N^{(\sigma_i)}$. The possible macrostates are 
therefore characterised by the internal energy $E$ and the two numbers $N^{(1)}$ and $N^{(2)}$ 
of spins in the state $1$ and $2$, respectively. 
In this two-dimensional plane the coordinates are chosen in such a way that 
the completely disordered macrostate $(N/3, N/3)$ where all spin states are equally likely corresponds to the   
magnetisation $(M_1=0, M_2=0)$. The (intensive) magnetisation $(m_1, m_2)$ with 
$m_l = M_l/N$ is related to the occupation numbers by the map 
\begin{eqnarray}
	m_1 & = & 1 - \frac{3}{N}\frac{N^{(1)}+N^{(2)}}{2} \\
	m_2 & = & \frac{\sqrt{3}}{2N}\left(N^{(2)} - N^{(1)}\right) \;.
\end{eqnarray}
The magnetisation of the Potts model with three spin states lies within an equilateral 
triangle in the $m_1m_2$ plane with the vertices at the points $(1, 0)$, $(-1/2, \sqrt{3}/2)$, and 
$(-1/2, -\sqrt{3}/2)$. These points correspond to the magnetisations of the three 
equivalent ground states of the model.  
The directions defined by the magnetisation vectors of these ground states 
(compare (\ref{eq:mag_richtung})) define the symmetry lines of the 
equilateral triangle. This triangle is depicted in Fig. \ref{pottsdreieck}.

The Hamiltonian (\ref{eq:potts_hamilton}) is invariant under 
the permutation group $S_3$ which is isomorphic to the group $C_{3v}$. 
The invariance group of the Potts model  
induces a set of linear mappings acting on the two magnetisation components $m_1$ and $m_2$. This 
two-dimensional representation 
of the group $C_{3v}$ is the irreducible representation commonly denoted by $\Gamma_3$. Note that this 
representation is also faithful. The entropy surface has therefore the
symmetry property (see Fig. \ref{pottsdreieck}) 
\begin{equation}
	\label{eq:potts_entro_sym}
	s(e, \Gamma_3(C_{3v})(m_1, m_2)) = s(e, m_1, m_2) \;.
\end{equation}

The appearance of the ground state magnetisation at the vertices of an equilateral triangle 
suggests that  
the extrema of the entropy as a function of the two magnetisation 
components for a fixed energy appear either at zero magnetisation or for non-zero 
magnetisations along the symmetry directions of the triangle specified by the angles 
$0$, $2\pi/3$, and $4\pi/3$. Here the angles are defined with respect to the $m_1$ axis on 
which the vertex $(1,0)$ of the equilateral triangle lies.

\begin{figure}%[htb]
\begin{center}
\epsfig{file=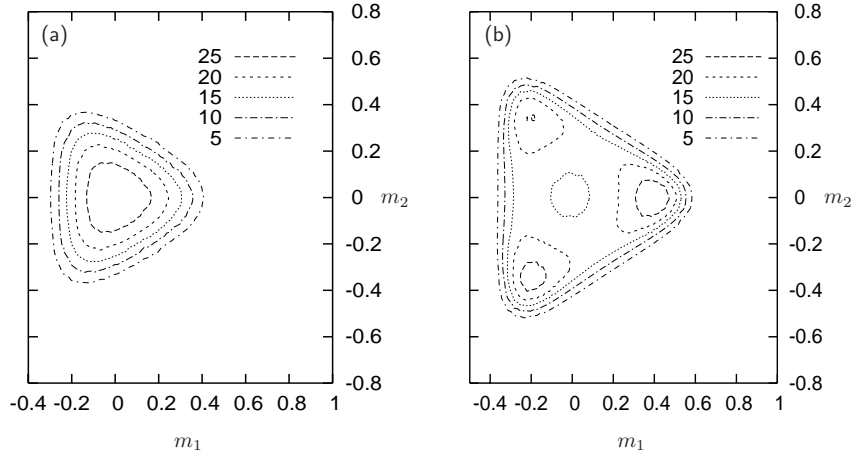,width=20em, angle=270}
\end{center}
\caption{\label{bildhoehen} \small Level curves of the density of states of the Potts 
model with $L=12$ for the energies $-1.166$ (a) and $-1.264$ 
(b). Three equivalent 
maxima appear along the symmetry lines of the equilateral triangle defined by the ground state magnetisations 
if the energy is below the critical value $-1.206$. 
Note that the density of states exhibits small asymmetries inevitable in Monte Carlo calculations. 
}
\end{figure}

\begin{figure}%[htb]
\begin{center}
\epsfig{file=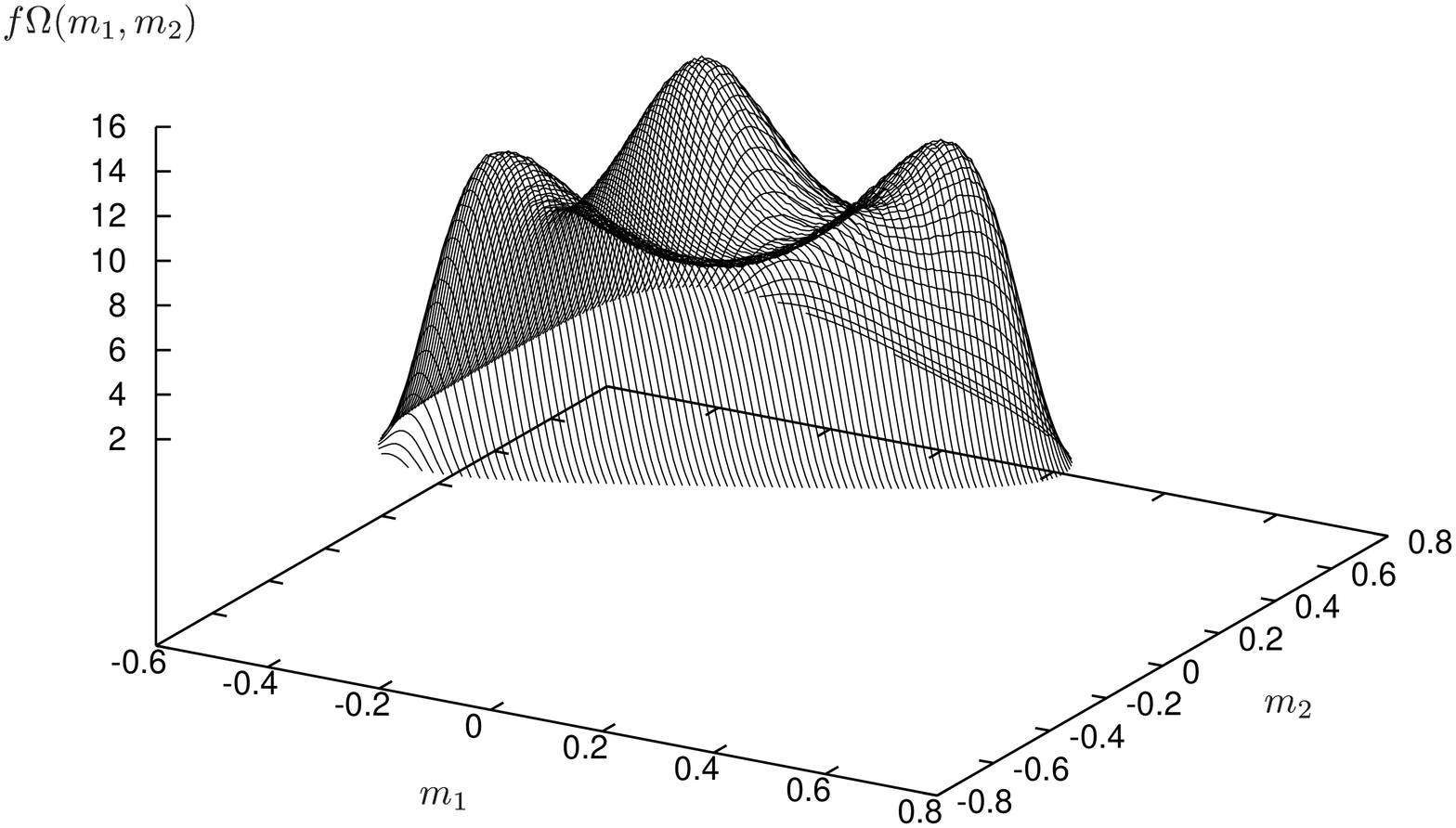,width=16em, angle=270}
\end{center}
\caption{\label{bildomega} \small The density of states of the Potts model with $L=12$ for the energy $e=-1.264$ 
below the critical point. The density of states is shown as the maxima are more pronounced in the density of state than in the 
microcanonical entropy. The surface shows three pronounced maxima and
reveals the three-fold symmetry of the entropy surface (compare also Fig.
3.25 in \cite{gros01}). The simulation gives the entropy up to a trivial additive constant. 
This has the consequence that the depicted density of states is only determined 
up to a multiplicative factor $f$.} 
\end{figure}

The entropy surface of the Potts model in two dimensions with finite linear extension $L$ 
for a fixed energy above the 
critical point $e_c$ at which the transition to a non-zero spontaneous magnetisation takes place exhibits a single maximum 
at zero magnetisation. Below $e_c$ three equivalent maxima show up for non-zero magnetisations. 
They appear along the bisectors of the equilateral triangle within which lie all possible macrostates 
of the Potts model for fixed energy. The extremum at zero magnetisation corresponds to a 
minimum of the entropy for energies below the critical point. 
This behaviour can be illustrated with a Potts system of linear extension $L=12$. 
The critical energy of the finite system with 
$144$ Potts spins is $-1.206\pm 0.003$.
Fig. \ref{bildhoehen} shows the level curves of the density 
of states of the Potts model for both an energy above 
and below the critical value $e_c$.
The density of states as a two-dimensional manifold for the energy below $e_c$ 
is shown in Fig. \ref{bildomega}. Note that 
the maxima and the minima of the density of states appear at the same magnetisations as the extrema of the 
entropy as the logarithm is a monotonic function.

Consider the maximum that shows up along the $m_1$ direction in the order parameter space. The corresponding 
stable low energy phase has the invariance group $G = \{ E, \sigma^{(1)}\} \subset C_{3v}$. Here $\sigma^{(1)}$ 
denotes the reflection about the $m_1$ direction, similarly $\sigma^{(2)}$ and $\sigma^{(3)}$ denote the 
reflection about the symmetry lines defined by the angles $2\pi/3$ and $4\pi/3$. The group 
$C_{3v}$ can be decomposed into left cosets with respect to $G$ giving the additional 
cosets $C_3G = \{C_3, \sigma^{(3)}\}$ and $C_3^2G = \{C_3^2, \sigma^{(2)}\}$. The transformation 
$C_3$ applied onto the $m_1$ direction gives the direction $2\pi/3$. The corresponding phase has the 
symmetry group $G^{(2)}= C_3GC_3^{-1}= \{E, \sigma^{(2)}\}$ (see relations 
(\ref{eq:transformiertesym}) and (\ref{eq:konjugiertesym})).

The three-state Potts model can be investigated analytically within the
plaquette approximation \cite{gros01}. This approach also reveals the
three-fold $C_{3v}$ symmetry of the entropy surface for fixed energies.

\subsection{Four-state vector Potts model}

An other possible generalisation of the Ising model is the $xy$ model \cite{Kogu79} specified by the 
Hamiltonian 
\begin{equation}
	\label{eq:xy_hamilton}
	\mathcal{H} = -\sum_{\left< i,j \right>}
	\vec{\sigma}_i \cdot \vec{\sigma}_j = 
	-\sum_{\left< i,j \right>} \cos(\varphi_i-\varphi_j)
\end{equation}
with the two-dimensional unit vector $\vec{\sigma}_i \in S^1$. The angle $\varphi_i$ 
characterises the direction of the vector $\vec{\sigma}_i$ in the plane. The magnetisation operator 
is given by 
\begin{equation}
	\vec{\mathcal{M}} = \sum_i \vec{\sigma}_i\;.
\end{equation}
The possible intensive magnetisations of the model for a fixed energy lie within the 
unit circle of the $xy$ plane. 
The invariance group $O(2)$ of the Hamiltonian induces 
the symmetry property 
\begin{equation}
	s(e, O(2)(\vec{m})) = s(e, \vec{m})
\end{equation}
of the entropy which has the consequence that the entropy depends only on the modulus 
$|\vec{m}|$ of the magnetisation vector $\vec{m}$. 

The continuous $xy$ model can be discretised by restricting the angle $\varphi_i$ to the values 
$2\pi k_i/q$ with $k_i = 0, \ldots, q-1$. The resulting model is often called $q$-state vector 
Potts model or $q$-state clock model.    
The Hamilton function of the four-state vector Potts model is 
\begin{equation}
	\label{eq:z4_hamilton}
  \mathcal{H} = -\sum_{\left<i,j\right>} \cos\left(\frac{\pi}{2}(k_i-k_j)\right) \;,
\end{equation}
where the spins $k_i$ can take on the values 0, 1, 2, and 3 and may be visualised by unit vectors with angles 
$0$, $\pi/2$, $\pi$, and $3\pi/2$ in a two-dimensional plane. The system has four equivalent 
ferromagnetically ordered ground states. The possible extensive energies $E$ of 
the spin configurations $k$ of the system 
are even numbers in the interval $[-2N, 2N]$. The magnetisation $(M_1, M_2)$ is given by 
\begin{eqnarray}
  M_1 & = & \sum_i \cos\left( \frac{\pi}{2} k_i\right) \\
  M_2 & = & \sum_i \sin\left( \frac{\pi}{2} k_i\right) \;. 
\end{eqnarray}
With the total numbers $N^{(k_i)}$ of spins in the spin state $k_i$ of a given configuration $k$ 
the intensive magnetisation can be expressed as 
\begin{eqnarray}
  m_1 & = & \frac{1}{N} (N^{(0)} - N^{(2)}) \\
  m_2 & = & \frac{1}{N} (N^{(1)} - N^{(3)}) \;.
\end{eqnarray}
The four equivalent ground states have the specific magnetisations $(1,0)$, $(0,1)$, $(-1,0)$, and 
$(0,-1)$. These points define a square in the magnetisation plane containing all possible macrostates of the 
four-state vector Potts model for a fixed energy. 

The Hamiltonian (\ref{eq:z4_hamilton}) is invariant under those transformations of the 
group $O(2)$ that leave a square in the magnetisation space invariant. The corresponding 
group is the group $C_{4v}$ with the rotations $C_4$, $C_4^2$, and $C_4^3$ about the angles 
$\pi/2$, $\pi$, and $3\pi/2$. It also contains the reflections $\sigma^{(1)}$ and 
$\sigma^{(2)}$ about the $m_2$ and $m_1$ direction and the reflections $\sigma^{(u)}$ about the 
direction $m_1=m_2$ and $\sigma^{(v)}$ about the direction $m_1=-m_2$. The representation of the 
group $C_{4v}$ that shows up as the symmetry operations on the magnetisation of the physical system 
is the two-dimensional, irreducible representation $\Gamma_5$. 
Apart from $\Gamma_5$ the group $C_{4v}$ has 
four additional one-dimensional irreducible representations.      
The appearance of the spontaneous magnetisation of the ground states on the 
lines $m_1=0$ and $m_2=0$ suggests that the non-zero spontaneous magnetisations of the system with higher energies 
are along the same directions.

\begin{figure}%[htb]
\begin{center}
\epsfig{file=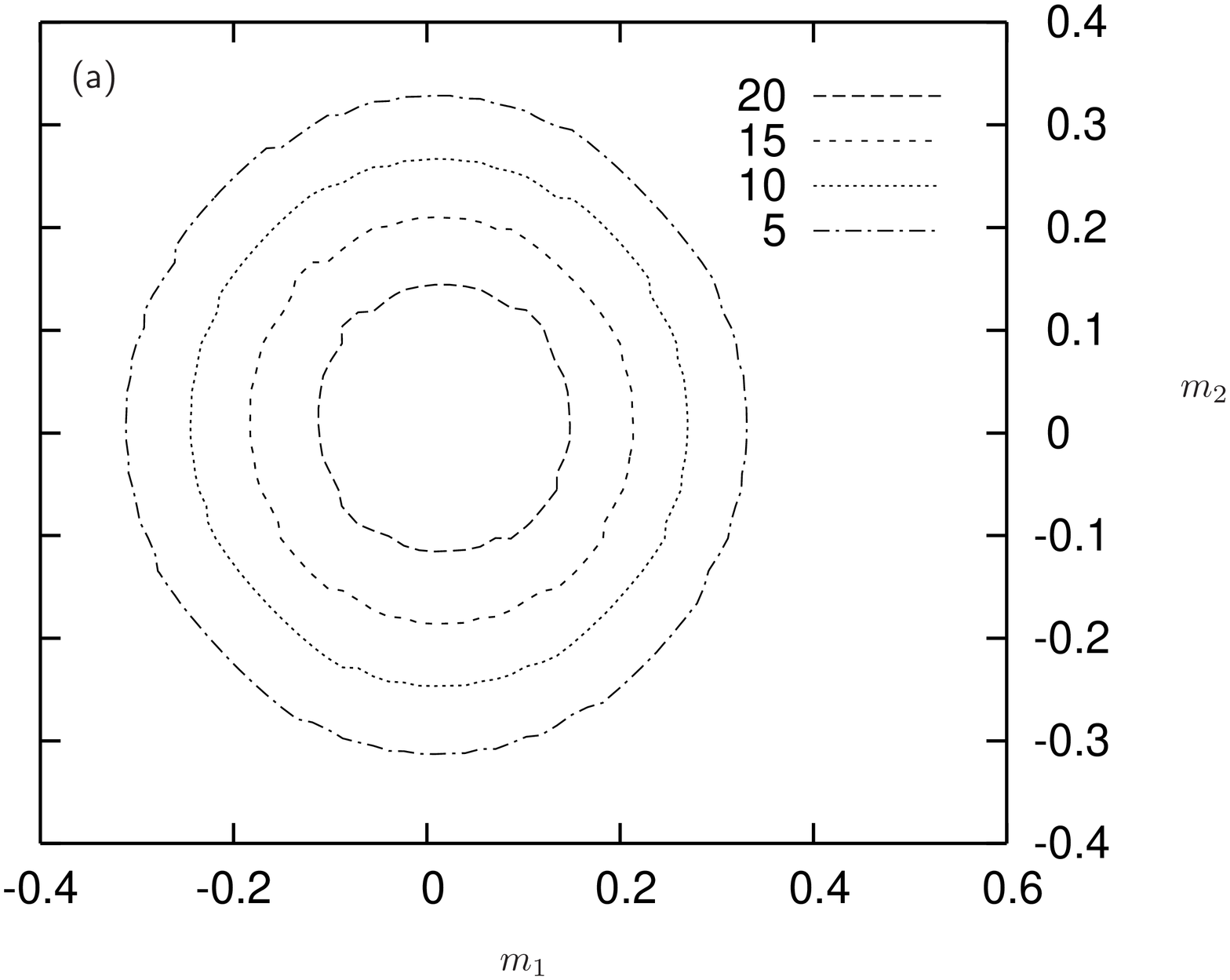,width=15em, angle=270}
\epsfig{file=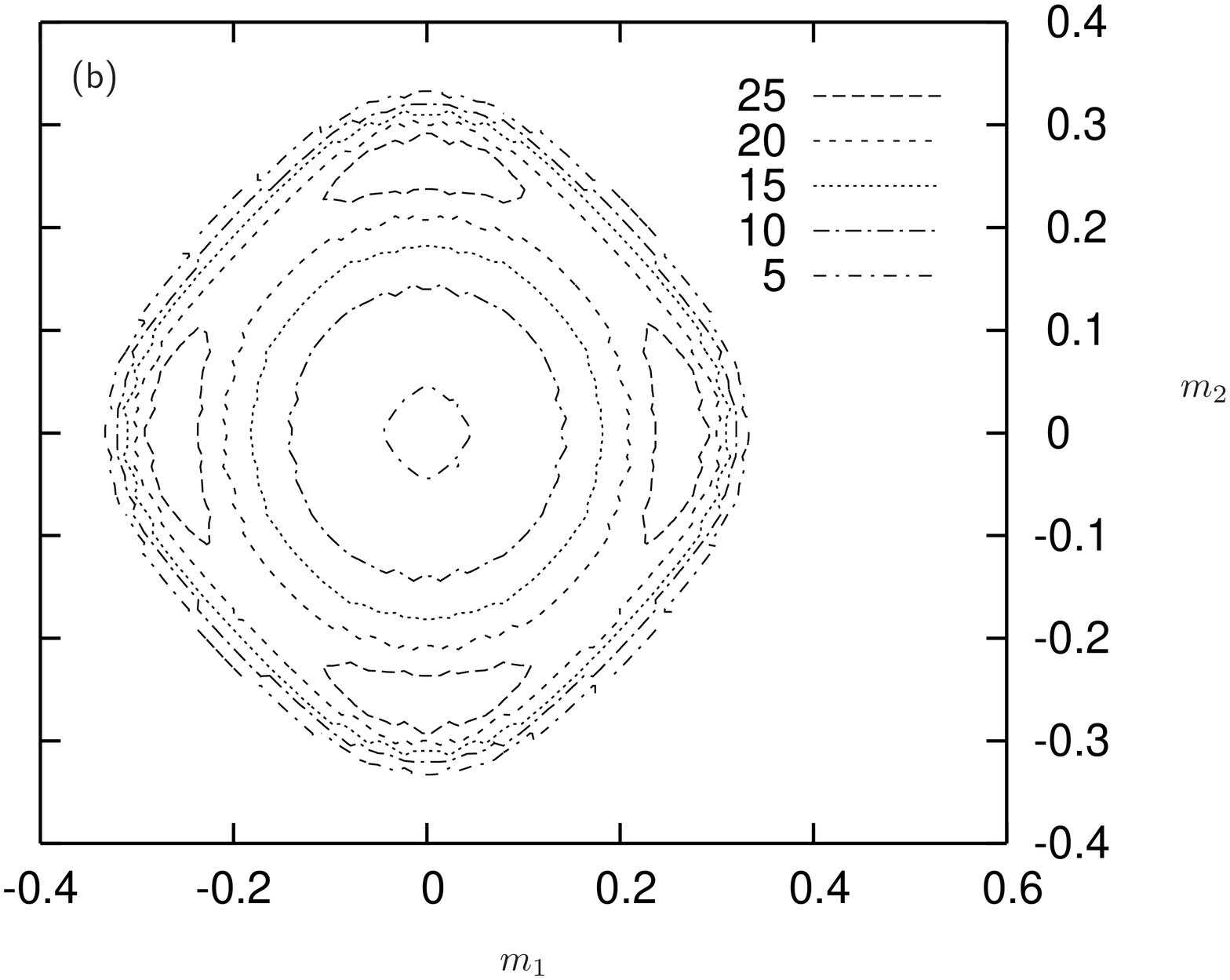,width=15em, angle=270}
\end{center}
\caption{\label{bildhoehen_z4} \small Level curves of the density of states of the four-states vector Potts 
model in two dimensions with linear extension $L=8$ for the energies $-0.5$ (a) and $-0.968$ 
(b). The entropy has only one maximum at zero magnetisation for energies above
the critical value $-0.7$. A minimum at zero magnetisation and four equivalent 
maxima appear along the bisectors of the square defined by the ground state magnetisations 
if the energy is below the critical value $-0.7$. 
}
\end{figure}

The four-state vector Potts model with $64$ spins in two dimensions undergoes a second order phase transition 
at the energy $e_c=-0.7\pm 0.005$. In Fig. \ref{bildhoehen_z4} the contour plots for entropies above and below 
the critical energy are shown. %The density of states for an energy below this value 
%is show in Fig. \ref{bildomega_z4}. 
The system develops four equivalent maxima along the lines $m_i=0$ if the 
energy is below the critical value. Only one maximum at zero magnetisation is present for energies above $e_c$. 
Saddle points of the entropy surface 
turn up along the second type of symmetry lines 
$m_1=\pm m_2$ for energies below $e_c$.% (see Fig. \ref{bildoentrschnitt_z4}). 
The high energy 
symmetry $C_{4v}$ of the four-state vector Potts model is broken down to the low energy symmetry group 
comprising the identity and the reflection about the direction $\mu^{(0)}$ defined by the 
equilibrium magnetisation through the decomposition (\ref{eq:mag_richtung}).    

In this subsection only the four-state vector Potts model as an example of a
system with four degrees of freedom is investigated. A comparison to a
different system with four degrees of freedom such as the ordinary Potts
model seems to be desirable. In particular the behaviour in the low energy
regime is of special interest but is left to future studies.

\section{Conclusion}

In this paper it has been shown that the isospin symmetry $G_S$ of the
Hamiltonian of a spin system determines the symmetry properties of the microcanonical entropy 
surface of a finite system at constant energies. The invariance group $G_S$
acts thereby through the irreducible representation $D(G_S)$ on the magnetisation 
components $m_l$. At the magnetisations $\{D(g_S)(m)| g_S \in G_S\}$ the
entropy have the same value.
The microcanonical order parameter below the critical point is a vector in the order parameter 
space with a finite modulus and defines therefore a direction in the order parameter space 
for the low energy phase. The 
invariance group $D(G)$ of this direction is a proper subgroup of $D(G_S)$ reflecting the spontaneous breakdown of the 
symmetry of the equilibrium macrostate of the system in the low energy
phase. This has been demonstrated for various spin models. 
The symmetry group of a spin model on the microscopical level leads already to a 
qualitative grasp of the microcanonical entropy surface of finite spin
systems. This qualitative picture of the microcanonical entropy surface
allows for example to scrutinise the consistency of estimates of the entropy that are
obtained by numerical simulations. Furthermore the knowledge of exact
symmetries of the entropy surface may be used to symmetrise the numerical
results and thereby improve the statistics of the simulation. The fact that
the various maxima of the entropy surface are physically identical can be
used to restrict
the investigation of the behaviour of the entropy function in the vicinity
of the equilibrium macrostate to the analysis of one of the equivalent maxima.

\ack

The author would like to thank Alfred H\"uller and Michel Pleimling for stimulating 
discussions and Efim Kats for helpful comments on the manuscript.

\Figures

\end{document}